\documentclass[a4paper,12pt]{article}%
\usepackage{amsmath}
\usepackage{amsfonts}
\usepackage{amssymb}
\begin{document}

\title{\textsf{Lost equivalence of nonlinear sigma and $CP^{1}$ models on
noncommutative space}}
\author{\textsf{Hideharu Otsu } \thanks{otsu@vega.aichi-u.ac.jp}\\Faculty of Economics, \\Aichi University, Toyohashi, Aichi 441-8522, Japan
\and \textsf{Toshiro Sato} \thanks{tsato@matsusaka-u.ac.jp}\\Faculty of Policy Science, Matsusaka University, \\Matsusaka, Mie 515-8511, Japan
\and \textsf{Hitoshi Ikemori } \thanks{ikemori@biwako.shiga-u.ac.jp}\\Faculty of Economics, Shiga University, \\Hikone, Shiga 522-8522, Japan
\and \textsf{Shinsaku Kitakado }\thanks{kitakado@ccmfs.meijo-u.ac.jp}\\Department of Information Sciences, Meijo University, \\Tempaku, Nagoya 486-8502, Japan }
\date{}
\maketitle

\begin{abstract}
We show that the equivalence of nonlinear sigma and $CP^{1}$ models which is
valid on the commutative space is broken on the noncommutative space. This
conclusion is arrived at through investigation of new BPS solitons that do not
exist in the commutative limit.

\end{abstract}

\newpage

\section{Introduction}

Field theories on the noncommutative space have been extensively studied in
the last few years. Particularly, BPS solitons are worth examined, because
they might not share the common features with those on the commutative space
\cite{Harvey:2001yn}\cite{Nekrasov:1998ss}\cite{Gopakumar:2000zd}.

Solitons in the $CP^{1}$ model on two dimensional noncommutative space have
been studied in \cite{Lee:2000ey} and further developed in
\cite{Furuta:2002ty} in connection with the dynamical aspects of the theory.
The non-BPS solitons, that do not exist in the commutative case, have been
studied in \cite{Furuta:2002nv}. In the previous paper \cite{Otsu:2003fq}, we
have reported on a set of new BPS solitons in the noncommutative $CP^{1}$
model, that does not exist in the commutative limit. On the other hand,
solitons in the $U(n)$ nonlinear sigma model have been studied on the
noncommutative space \cite{Lechtenfeld:2001uq}\cite{Lechtenfeld:2001aw}%
\cite{Lechtenfeld:2001gf}. 

In this paper, we investigate the new aspects one encounters when the two
dimensional space is promoted to the noncommutative space. We consider the
nonlinear sigma model, which has been discussed as the modified $U(2)$ sigma
model in \cite{Lechtenfeld:2001uq}\cite{Lechtenfeld:2001aw}, and $CP^{1}$
model defined on the noncommutative space $\mathbb{R}_{NC}^{2}$ with
commutative time. Although our discussions are concerned with the static
solitons on $\mathbb{R}_{NC}^{2}\times\mathbb{R}$, these solutions can also be
considered as the instanton solutions on $\mathbb{R}_{NC}^{2}$.

On the commutative space $\mathbb{R}^{2}$, there exists a definite
correspondence between the configurations of the nonlinear sigma model and the
configurations of the $CP^{1}$ model and both models are equivalent. We shall
see, however, that on the noncommutative space such a correspondence is
destroyed. In fact, there exists a BPS soliton in the nonlinear sigma model
that does not have the counterpart in the $CP^{1}$ model. Furthermore, the
correspondence relation between the two models is different for solitons and
anti-solitons. A new BPS anti-soliton solution in the $CP^{1}$ model has been
found on the noncommutative space that does not exist in the commutative space.

In what follows, after a brief description of the $CP^{1}$ model and the
nonlinear sigma model on the commutative space in the section 2, we proceed,
in section 3, to the description of these models and their BPS\ solitons on
the noncommutative space. In section 4 we investigate the subtle properties of
the BPS\ solitons discussed in section 3. Section 5 is devoted to summary.

\section{Models on the commutative space\quad}

In this section we shall describe the $CP^{1}$ model and the nonlinear sigma
model on the commutative space. We focus our attention to the topological
properties of the configurations and their interrelations in both models.

The field variable in $CP^{1}$ model is given by
\begin{equation}
\Phi=\left(
\begin{array}
[c]{l}%
\phi_{1}\\
\phi_{2}%
\end{array}
\right)
\end{equation}
with the constraint
\begin{equation}
\Phi^{\dagger}\Phi=1,
\end{equation}
which implies that $\Phi$ spans the $S^{3}$ space. The lagrangian is written
as
\begin{equation}
L\mathbf{=}\int d^{2}x(\left\vert D_{t}\Phi\right\vert ^{2}-\sum
\limits_{i=1}^{2}\left\vert D_{i}\Phi\right\vert ^{2}),\label{Lccp}%
\end{equation}
where $D_{a}$ is a covariant derivative defined as%
\begin{align}
D_{a}\Phi &  =\partial_{a}\Phi-i\Phi A_{a},\nonumber\\
A_{a} &  =-i\Phi^{\dagger}\partial_{a}\Phi.
\end{align}
This system has a local $U(1)$\ symmetry under the transformation
$\Phi\rightarrow e^{i\alpha}\Phi$. Consequently, the static configuration is
characterized by the homotopy class $\Pi_{2}\left(  S^{3}/S^{1}\right)  =Z$ .
The corresponding topological charge can be written as
\begin{align}
Q &  =\frac{-i}{2\pi}\int d^{2}x\epsilon_{ij}(D_{i}\Phi^{\dagger}D_{j}%
\Phi)\nonumber\\
&  =\frac{-i}{2\pi}\int d^{2}x\epsilon_{ij}(\partial_{i}\Phi^{\dagger}%
\partial_{j}\Phi).\label{Qccp}%
\end{align}
Although, the second equality is valid on the commutative space, it is not the
case on the noncommutative space. In section 3, we shall define the
topological charge through the first covariant form \cite{Lee:2000ey}%
\cite{Foda:2002nt}. We have the following energy bound for the static configuration,%

\begin{equation}
E=\int d^{2}x\sum\limits_{i=1}^{2}\left\vert D_{i}\Phi\right\vert ^{2}\geq
8\pi\left\vert Q\right\vert ,
\end{equation}
and the equality is satisfied for the BPS soliton (anti-soliton).

Next we recapitulate the notations of the nonlinear sigma model that is in
fact equivalent to the $CP^{1}$ model on the commutative space. Using the
variable $n^{a}$ with the constraint $\sum_{a=1}^{3}(n^{a})^{2}=1$, the
lagrangian can be written as
\begin{equation}
L=\int d^{2}x\left[  (\partial_{t}n^{a})^{2}-(\partial_{i}n^{a})^{2}\right]  ,
\label{Lcnls}%
\end{equation}
and the topological charge is expressed as
\begin{equation}
Q=\frac{1}{8\pi}\int d^{2}x\epsilon_{ij}\epsilon_{abc}n^{a}\partial_{i}%
n^{b}\partial_{j}n^{c}. \label{Qcnls}%
\end{equation}
Relation with the $CP^{1}$ variable $\Phi$ is
\begin{equation}
n^{a}=\Phi^{\dagger}\sigma^{a}\Phi, \label{hopf}%
\end{equation}
which leads to the equalities of (\ref{Lccp}) with (\ref{Lcnls}) and of
(\ref{Qccp}) with (\ref{Qcnls}) .

Using the projector $P\equiv\Phi\Phi^{\dagger}$ $\left(  P^{2}=P\right)  $, we
can express the nonlinear sigma model in terms of the variable $U$
\begin{equation}
U\equiv2P-1
\end{equation}
which satisfies%
\begin{equation}
U^{2}=1.
\end{equation}
\newline Lagrangian, energy of static configuration, topological charge are
rewritten as
\begin{equation}
L=\frac{1}{2}\int d^{2}x\mathrm{tr}\left[  (\partial_{t}U)^{2}-2\partial
_{\bar{z}}U\partial_{z}U\right]  ,
\end{equation}%
\begin{equation}
E=\int d^{2}x\mathrm{tr}(\partial_{\bar{z}}U\partial_{z}U)
\end{equation}
and%
\begin{equation}
Q=\frac{1}{16\pi}\int d^{2}x\mathrm{tr}\left[  U\left(  \partial_{\bar{z}%
}U\partial_{z}U-\partial_{z}U\partial_{\bar{z}}U\right)  \right]  ,
\label{QcnlsU}%
\end{equation}
respectively. Here \textquotedblleft tr" denotes the trace of $2\times
2$\ matrices. The static configurations of nonlinear sigma model satisfy
$\sum_{a=1}^{3}(n^{a})^{2}=1$ or equivalently $U^{2}=1$. Thus they are
classified in terms of the homotopy class $\Pi_{2}\left(  S^{2}\right)  =Z$
and the corresponding topological charges are (\ref{Qcnls}) and (\ref{QcnlsU})
. The expressions using the variables $n^{a}$ and those using $U$ are
equivalent on the commutative space, where we have the relation%
\begin{equation}
U=n^{a}\sigma^{a}.
\end{equation}
However, this is not the case on the noncommutative space. In section 3 we
shall extend the nonlinear sigma model to the noncommutative space using the
lagrangian, topological charge written in terms of $U$ .

Finally, we note in passing that the configuration in $CP^{1}$ model and the
configuration in the nonlinear sigma model can be related with each other
(\ref{hopf}). This relation can also be solved for $\Phi$
\begin{equation}
\Phi=\frac{1}{\sqrt{2}}\frac{e^{i\alpha}}{\sqrt{1-n^{3}}}\binom{n^{1}+in^{2}%
}{1-n^{3}}. \label{ccp-n}%
\end{equation}
Furthermore, if the use is made of the relation (\ref{ccp-n}), the lagrangians
are also equivalent. Consequently, on the commutative space, the nonlinear
sigma model and the $CP^{1}$ model are equivalent.

\section{Models on the noncommutative space}

In this section we shall investigate the $CP^{1}$ and the nonlinear sigma
models on the noncommutative space $\mathbb{R}_{NC}^{2}$ \cite{Lee:2000ey}%
\cite{Furuta:2002ty}\cite{Furuta:2002nv}\cite{Otsu:2003fq}\cite{Ghosh:2004ee}%
\cite{Ghosh:2003cu}\cite{Ghosh:2003ka}\cite{Murugan:2002rz}. The space
coordinates obey the commutation relation%
\begin{equation}
\left[  x,y\right]  =i\theta
\end{equation}
or%
\begin{equation}
\left[  z,\bar{z}\right]  =\theta>0,
\end{equation}
when written in the complex variables, $z=\frac{1}{\sqrt{2}}(x+iy)$ and
$\bar{z}=\frac{1}{\sqrt{2}}(x-iy)$. The Hilbert space can be described in
terms of the energy eigenstates $\left\vert n\right\rangle $ of the harmonic
oscillator whose creation and annihilation operators are $\bar{z}$ and $z$
respectively,%
\begin{align}
z\left\vert n\right\rangle  &  =\sqrt{\theta n}\left\vert n-1\right\rangle ,\\
\bar{z}\left\vert n\right\rangle  &  =\sqrt{\theta(n+1)}\left\vert
n+1\right\rangle .\nonumber
\end{align}
Space integrals on the commutative space are replaced by the trace on the
Hilbert space
\begin{equation}
\int d^{2}x\Rightarrow\mathrm{Tr}_{\mathcal{H}}\mathrm{,}%
\end{equation}
where, $\mathrm{Tr}_{\mathcal{H}}$ denotes the trace over the Hilbert space
as
\begin{equation}
\text{$\mathrm{Tr}_{\mathcal{H}}$}\mathcal{O}=2\pi\theta\sum_{n=0}^{\infty
}\left\langle n\right\vert \mathcal{O}\left\vert n\right\rangle .
\end{equation}
The derivatives with respect to $z$ and $\bar{z}$ are defined by $\partial
_{z}=-\theta^{-1}\left[  \bar{z},\right]  $\ and $\partial_{\bar{z}}%
=\theta^{-1}\left[  z,\right]  $.

The $CP^{1}$ lagrangian is%
\begin{equation}
L=\mathrm{Tr}_{\mathcal{H}}(|D_{t}\Phi|^{2}-|D_{z}\Phi|^{2}-|D_{\bar{z}}%
\Phi|^{2}),
\end{equation}
where $\Phi$ is a 2-component complex vector with the constraint
$\Phi^{\dagger}\Phi=1$. The covariant derivative is defined by%
\begin{equation}
D_{a}\Phi=\partial_{a}\Phi-i\Phi A_{a},\ A_{a}=-i\Phi^{\dagger}\partial
_{a}\Phi.
\end{equation}
For the static configuration, topological charge and energy are given by%

\begin{equation}
Q=\frac{-i}{2\pi}\mathrm{Tr}_{\mathcal{H}}(\epsilon_{ij}D_{i}\Phi^{\dagger
}D_{j}\Phi)=\frac{1}{2\pi}\mathrm{Tr}_{\mathcal{H}}\left(  \left\vert
D_{z}\Phi\right\vert ^{2}-\left\vert D_{\bar{z}}\Phi\right\vert ^{2}\right)
\end{equation}
and%
\begin{equation}
E=\mathrm{Tr}_{\mathcal{H}}\left(  \left\vert D_{z}\Phi\right\vert
^{2}+\left\vert D_{\bar{z}}\Phi\right\vert ^{2}\right)  \geq2\pi\left\vert
Q\right\vert .
\end{equation}
The configuration which saturates the energy bound satisfies the BPS soliton
equation
\begin{equation}
D_{\bar{z}}\Phi=(1-\Phi\Phi^{\dagger})z\Phi=0
\end{equation}
or BPS anti-soliton equation\smallskip%
\begin{equation}
D_{z}\Phi=(1-\Phi\Phi^{\dagger})\bar{z}\Phi=0.
\end{equation}

The following BPS soliton (anti-soliton) solutions are known. The solutions
that have the counterparts in the $CP^{1}$\ model on the commutative space
\cite{Lee:2000ey} are soliton solutions
\begin{equation}
W=\left(
\begin{array}
[c]{c}%
z^{n}\\
1
\end{array}
\right)  \label{sLee}%
\end{equation}
with $Q=n,E=2\pi n$ and anti-soliton solutions%
\begin{equation}
W=\left(
\begin{array}
[c]{c}%
\bar{z}^{n}\\
1
\end{array}
\right)  \label{asLee}%
\end{equation}
with $Q=-n,E=2\pi n$. Here%
\begin{equation}
\Phi=W\frac{1}{\sqrt{W^{\dagger}W}}\ .
\end{equation}
(\ref{sLee}) and (\ref{asLee}) when expressed in terms of $P=\Phi\Phi
^{\dagger}$ are respectively%
\begin{equation}
P=\left(
\begin{array}
[c]{cc}%
z^{n}\dfrac{1}{\bar{z}^{n}z^{n}+1}\bar{z}^{n} & z^{n}\dfrac{1}{\bar{z}%
^{n}z^{n}+1}\\
\dfrac{1}{\bar{z}^{n}z^{n}+1}\bar{z}^{n} & \dfrac{1}{\bar{z}^{n}z^{n}+1}%
\end{array}
\right)  \label{sLee-P}%
\end{equation}
and%

\begin{equation}
P=\left(
\begin{array}
[c]{cc}%
\bar{z}^{n}\dfrac{1}{z^{n}\bar{z}^{n}+1}z^{n} & \bar{z}^{n}\dfrac{1}{z^{n}%
\bar{z}^{n}+1}\\
\dfrac{1}{z^{n}\bar{z}^{n}+1}z^{n} & \dfrac{1}{z^{n}\bar{z}^{n}+1}%
\end{array}
\right)  . \label{asLee-P}%
\end{equation}
Furthermore, there are solutions that do not exist on the commutative space
\cite{Otsu:2003fq}, they are $Q=n,E=2\pi n$ soliton solutions;%
\begin{equation}
{\Phi}=\left(
\begin{array}
[c]{c}%
\dfrac{1}{\sqrt{\prod_{k=1}^{n}(\bar{z}z+k\theta)}}z^{n}\\
\sum_{m=0}^{n-1}\left\vert m\right\rangle \left\langle m\right\vert
\end{array}
\right)  \label{sOSIKcp}%
\end{equation}
and $Q=-n,E=2\pi n$ anti-soliton solutions;%
\begin{equation}
\Phi=\left(
\begin{array}
[c]{c}%
\bar{z}^{n}\dfrac{1}{\sqrt{\prod_{k=1}^{n}(\bar{z}z+k\theta)}}\\
0
\end{array}
\right)  . \label{asOSIKcp}%
\end{equation}
(\ref{sOSIKcp}) (\ref{asOSIKcp}) when expressed in terms of $P=\Phi
\Phi^{\dagger}$ are respectively%
\begin{equation}
P=\left(
\begin{array}
[c]{cc}%
1 & 0\\
0 & \sum_{m=0}^{n-1}\left\vert m\right\rangle \left\langle m\right\vert
\end{array}
\right)  \label{sOSIK-P}%
\end{equation}
and%

\begin{equation}
P=\left(
\begin{array}
[c]{cc}%
1-\sum_{m=0}^{n-1}\left\vert m\right\rangle \left\langle m\right\vert  & 0\\
0 & 0
\end{array}
\right)  . \label{asOSIK-P}%
\end{equation}

Next we turn to the nonlinear sigma model on the noncommutative space. We
start from the model on the commutative space expressed in terms of $U$.
Lagrangian and topological charge are respectively
\begin{align}
L  &  =\frac{1}{2}\mathrm{Tr}_{\mathcal{H}}\left[  \text{\textrm{tr}}\left(
(\partial_{t}U)^{2}-2\partial_{{\bar{z}}}U\partial_{z}U\right)  \right]
\nonumber\\
&  =\frac{1}{2}\mathrm{Tr}_{\mathcal{H}}\left[  \mathrm{tr}(\partial_{t}%
P)^{2}\right]  -\theta^{-2}\mathrm{Tr}_{\mathcal{H}}\left[  \mathrm{tr}\left(
[z,P][P,\bar{z}]\right)  \right]
\end{align}
and%
\begin{equation}
Q=\frac{1}{16\pi\theta^{2}}\mathrm{Tr}_{\mathcal{H}}\left[  \mathrm{tr}\left(
U\left(  [\bar{z},U][z,U]-[z,U][\bar{z},U]\right)  \right)  \right]  .
\end{equation}
Energy for the static configuration is expressed as%
\begin{align}
\theta^{2}E  &  =\mathrm{Tr}_{\mathcal{H}}\left[  \mathrm{tr}(\left[
z,P\right]  \left[  P,\bar{z}\right]  )\right] \nonumber\\
&  =\mathrm{Tr}_{\mathcal{H}}\left[  \mathrm{tr}\left(  P\left[  \bar
{z},P\right]  \left[  z,P\right]  -P\left[  z,P\right]  \left[  \bar
{z},P\right]  \right)  \right]  +\mathrm{Tr}_{\mathcal{H}}\left[
\mathrm{tr}\left(  F^{\dagger}F+FF^{\dagger}\right)  \right] \nonumber\\
&  \geq\mathrm{Tr}_{\mathcal{H}}\left[  \mathrm{tr}\left(  P\left[  \bar
{z},P\right]  \left[  z,P\right]  -P\left[  z,P\right]  \left[  \bar
{z},P\right]  \right)  \right] \nonumber\\
&  =2\pi\theta^{2}Q+\frac{1}{2}\mathrm{Tr}_{\mathcal{H}}\left[  \mathrm{tr}%
\left(  \left[  \bar{z},P\right]  \left[  z,P\right]  -\left[  z,P\right]
\left[  \bar{z},P\right]  \right)  \right]  , \label{Ebound}%
\end{align}
where $F=(1-P)zP$ \cite{Lechtenfeld:2001aw}\cite{Gopakumar:2001yw}.\quad The
second term on the last line of Eq. (\ref{Ebound}) is zero for the finite
energy configuration. Consequently, the energy bound%

\begin{equation}
E\geq2\pi Q
\end{equation}
is satisfied for $Q>0$. The BPS soliton equation \cite{Lechtenfeld:2001aw}%
\cite{Gopakumar:2001yw}\cite{Hadasz:2001cn} is%
\begin{equation}
(1-P)zP=0. \label{BPS-P}%
\end{equation}
Similarly, for $Q<0$ the BPS anti-soliton equation is%
\begin{equation}
(1-P)\bar{z}P=0. \label{aBPS-P}%
\end{equation}
With $P=\Phi\Phi^{\dagger}$, the BPS equations (\ref{BPS-P}) and
(\ref{aBPS-P}) are consistent with those of $CP^{1}$ model.

Finally, let us see the relation of configurations in the noncommutative
$CP^{1}$ model with those of nonlinear sigma model. The configuration of
nonlinear sigma model can be obtained from that of $CP^{1}$ model through
$U=2\Phi\Phi^{\dagger}-1\ $\ or $P=\Phi\Phi^{\dagger}$. Obtaining $CP^{1}$
configuration from that of nonlinear sigma model is not straightforward. In
what follows we shall see a concrete relation between the configurations of
both models considering as an example the BPS soliton (anti-soliton) configuration.

The $CP^{1}$ BPS soliton (\ref{sLee}) (\ref{sOSIKcp}) (anti-soliton
(\ref{asLee}) (\ref{asOSIKcp})) are solutions of the nonlinear sigma model
through the relation $P=\Phi\Phi^{\dagger}$. Consider next a new BPS soliton
of the nonlinear sigma model expressed in terms of $P,$%

\begin{equation}
P=\left(
\begin{array}
[c]{cc}%
\sum_{m=0}^{k-1}\left\vert m\right\rangle \left\langle m\right\vert  & 0\\
0 & \sum_{m=0}^{n-1}\left\vert m\right\rangle \left\langle m\right\vert
\end{array}
\right)  , \label{newS}%
\end{equation}
which satisfies the BPS soliton equation (\ref{BPS-P}). Topological charge and
energy are $Q=k+n$ and $E=2\pi(k+n)$, respectively. If we require $P=\Phi
\Phi^{\dagger}$ for this configuration, we have%
\begin{equation}
\Phi^{\dagger}\Phi=\sum_{m=0}^{k+n-1}\left\vert m\right\rangle \left\langle
m\right\vert \neq1,
\end{equation}
which shows the absence of the corresponding $CP^{1}$ configuration. We note,
in this connection that $\Phi$ can be expressed as
\begin{equation}
\quad\quad\Phi=\left(
\begin{array}
[c]{c}%
\sum\limits_{m=0}^{k-1}\left\vert m\right\rangle \left\langle m+n\right\vert
\\
\sum\limits_{m=0}^{n-1}\left\vert m\right\rangle \left\langle m\right\vert
\end{array}
\right)  ,
\end{equation}
but $\Phi^{\dagger}\Phi=1$ is not valid for finite $k$ . Only for
$k\rightarrow\infty$ we have $\Phi^{\dagger}\Phi=1$ which leads to the
$CP^{1}$ soliton (\ref{sOSIKcp}) or (\ref{sOSIK-P}).

Furthermore, consider a BPS anti-soliton
\begin{equation}
P=\left(
\begin{array}
[c]{cc}%
1-\sum_{m=0}^{n-1}\left\vert m\right\rangle \left\langle m\right\vert  & 0\\
0 & 1-\sum_{m=0}^{k-1}\left\vert m\right\rangle \left\langle m\right\vert
\end{array}
\right)  , \label{newAS}%
\end{equation}
which satisfies BPS anti-soliton equation (\ref{aBPS-P}). Topological charge
and energy are $Q=-(k+n)$ and $E=2\pi(k+n)$, respectively. When $k$ is sent to
infinity, it reduces to the $CP^{1}$ anti-soliton (\ref{asOSIKcp}) or
(\ref{asOSIK-P}). If we require $P=\Phi\Phi^{\dagger}$ for the anti-soliton
(\ref{newAS}), we get for example
\begin{equation}
\quad\quad\Phi=\left(
\begin{array}
[c]{c}%
\sum\limits_{m=n}^{\infty}\left\vert m\right\rangle \left\langle
2(m-n)+1\right\vert \\
\sum\limits_{m=k}^{\infty}\left\vert m\right\rangle \left\langle
2(m-k)\right\vert
\end{array}
\right)  , \label{newAScp}%
\end{equation}
which satisfies $\Phi^{\dagger}\Phi=1$. In this case, although the $CP^{1}$
configuration does exist, it does not have the commutative limit due to the
fact that infinitely many \textquotedblleft dislocations" of the state prevent
us from approaching the continuous configuration.

From what we have seen, we may conclude that the correspondence between the
configurations of the $CP^{1}$ model and the nonlinear sigma model is
destroyed. Concretely, we have shown that the soliton which cannot exist in
the $CP^{1}$ model can be found in the nonlinear sigma model as a BPS soliton.
Furthermore, we have found a new BPS anti-soliton solution (\ref{newAScp}) of
$CP^{1}$ model which does not exist on the commutative space.

\section{Properties of soliton solutions}

In this section, we shall analyze the properties of the solitons discussed in
the previous sections.

\subsection{Anti-solitons in the $CP^{1}$\ model}

\bigskip First let us look for the general form of $CP^{1}$\ anti-soliton that
corresponds to the anti-soliton (\ref{newAS}) in the nonlinear sigma model. We
can solve
\begin{equation}
\Phi\Phi^{\dagger}=\left(
\begin{array}
[c]{cc}%
1-\sum_{m=0}^{n-1}\left\vert m\right\rangle \left\langle m\right\vert  & 0\\
0 & 1-\sum_{m=0}^{k-1}\left\vert m\right\rangle \left\langle m\right\vert
\end{array}
\right)
\end{equation}
as
\begin{equation}
\Phi=\left(
\begin{array}
[c]{c}%
\alpha\mu^{\dag}\\
\beta\nu^{\dag}%
\end{array}
\right)  . \label{newAScp-g}%
\end{equation}
Here $\alpha,\beta$ satisfy
\begin{equation}
\alpha^{\dag}\alpha=1,\ \beta^{\dag}\beta=1
\end{equation}
and $\mu,\nu$
\begin{align}
\mu^{\dag}\mu &  =1,\ \nu^{\dag}\nu=1,\nonumber\\
\mu^{\dag}\nu &  =0,\ \nu^{\dag}\mu=0, \label{munu1}%
\end{align}%
\begin{equation}
\mu\mu^{\dag}+\nu\nu^{\dag}=1, \label{munu2}%
\end{equation}
and consequently%
\begin{equation}
\Phi^{\dag}\Phi=1,
\end{equation}
thus we can confirm that $\Phi$ is a variable of $CP^{1}$\ model. Using this
$\Phi$ we have%
\begin{equation}
P=\Phi\Phi^{\dag}=\left(
\begin{array}
[c]{cc}%
\alpha\alpha^{\dag} & 0\\
0 & \beta\beta^{\dag}%
\end{array}
\right)  .
\end{equation}

Let us introduce the operators $S_{N}$ and $P_{N}\ $,
\begin{equation}
S_{N}\equiv\sum_{m=0}^{\infty}\left\vert m+N\right\rangle \left\langle
m\right\vert ,\ P_{N}\equiv\sum_{m=0}^{N-1}\left\vert m\right\rangle
\left\langle m\right\vert ,
\end{equation}
with the properties%
\begin{align}
S_{N}^{\dag}S_{N}  &  =1,\ S_{N}S_{N}^{\dag}=1-P_{N},\nonumber\\
P_{N}S_{N}  &  =0=S_{N}^{\dag}P_{N}.
\end{align}
We can express as%
\begin{equation}
\alpha=S_{N},\ \beta=S_{K},
\end{equation}
and write $P$ as
\begin{equation}
P=\left(
\begin{array}
[c]{cc}%
S_{N}S_{N}^{\dag} & 0\\
0 & S_{K}S_{K}^{\dag}%
\end{array}
\right)  =\left(
\begin{array}
[c]{cc}%
1-P_{N} & 0\\
0 & 1-P_{K}%
\end{array}
\right)  ,
\end{equation}
thus $P$ satisfies the anti-BPS equation (\ref{aBPS-P}). \ 

If we choose $\mu,\nu$ as%
\begin{equation}
\mu=\sum_{p=0}^{\infty}\left\vert 2p+1\right\rangle \left\langle p\right\vert
,\ \nu=\sum_{p=0}^{\infty}\left\vert 2p\right\rangle \left\langle p\right\vert
,
\end{equation}
we can show that
\begin{align}
\mu^{\dag}\mu &  =\sum_{p^{\prime}=0}^{\infty}\sum_{p=0}^{\infty}\left\vert
p^{\prime}\right\rangle \left\langle 2p^{\prime}+1\right\vert \left.
2p+1\right\rangle \left\langle p\right\vert =\sum_{p=0}^{\infty}\left\vert
p\right\rangle \left\langle p\right\vert =1,\nonumber\\
\nu^{\dag}\nu &  =\sum_{p^{\prime}=0}^{\infty}\sum_{p=0}^{\infty}\left\vert
p^{\prime}\right\rangle \left\langle 2p^{\prime}\right\vert \left.
2p\right\rangle \left\langle p\right\vert =\sum_{p=0}^{\infty}\left\vert
p\right\rangle \left\langle p\right\vert =1,\nonumber\\
\mu\mu^{\dag}  &  =\sum_{p^{\prime}=0}^{\infty}\sum_{p=0}^{\infty}\left\vert
2p^{\prime}+1\right\rangle \left\langle p^{\prime}\right\vert \left.
p\right\rangle \left\langle 2p+1\right\vert =\sum_{p=0}^{\infty}\left\vert
2p+1\right\rangle \left\langle 2p+1\right\vert ,\nonumber\\
\nu\nu^{\dag}  &  =\sum_{p^{\prime}=0}^{\infty}\sum_{p=0}^{\infty}\left\vert
2p^{\prime}\right\rangle \left\langle p^{\prime}\right\vert \left.
p\right\rangle \left\langle 2p\right\vert =\sum_{p=0}^{\infty}\left\vert
2p\right\rangle \left\langle 2p\right\vert ,\nonumber\\
\nu^{\dagger}\mu &  =\sum_{p^{\prime}=0}^{\infty}\sum_{p=0}^{\infty}\left\vert
p^{\prime}\right\rangle \left\langle 2p^{\prime}\right\vert \left.
2p+1\right\rangle \left\langle p\right\vert =0,
\end{align}
thus satisfying (\ref{munu1}) and (\ref{munu2}). This gives
\begin{equation}
\Phi=\left(
\begin{array}
[c]{c}%
S_{n}\mu^{\dag}\\
S_{k}\nu^{\dag}%
\end{array}
\right)  =\left(
\begin{array}
[c]{c}%
\sum\limits_{m=n}^{\infty}\left\vert m\right\rangle \left\langle
2(m-n)+1\right\vert \\
\sum\limits_{m=k}^{\infty}\left\vert m\right\rangle \left\langle
2(m-k)\right\vert
\end{array}
\right)
\end{equation}
which appeared in the previous section (\ref{newAScp}) .

It is interesting to note that $S_{N}$ and $P_{N}$\ can be used to express our
solutions found in the previous work \cite{Otsu:2003fq} for anti-soliton
(\ref{asOSIKcp}) and (\ref{asOSIK-P})%
\begin{equation}
\Phi=\left(
\begin{array}
[c]{c}%
S_{n}\\
0
\end{array}
\right)  ,\ P=\left(
\begin{array}
[c]{cc}%
S_{n}S_{n}^{\dag} & 0\\
0 & 0
\end{array}
\right)  =\left(
\begin{array}
[c]{cc}%
1-P_{n} & 0\\
0 & 0
\end{array}
\right)
\end{equation}
and for soliton (\ref{sOSIKcp}) and (\ref{sOSIK-P})%
\begin{equation}
\Phi=\left(
\begin{array}
[c]{c}%
S_{n}^{\dag}\\
P_{n}%
\end{array}
\right)  ,\ P=\left(
\begin{array}
[c]{cc}%
S_{n}^{\dag}S_{n} & S_{n}^{\dag}P_{n}\\
P_{n}S_{n} & P_{n}%
\end{array}
\right)  =\left(
\begin{array}
[c]{cc}%
1 & 0\\
0 & P_{n}%
\end{array}
\right)  .
\end{equation}
These expressions are in accord with the solution generating technique
appearing in \cite{Hashimoto:2000kq}\cite{Hamanaka:2000aq}%
\cite{Hamanaka:2001dr}.

\subsection{Classification of noncommutative solitons}

We shall show in this subsection that the values of $\mathrm{tr}P$ at the
boundary of the Hilbert space can be used in classifying the solitons
discussed in section 3. The trace of field variable $P$ at the boundary of the
Hilbert space is defined as
\begin{equation}
\left\langle \mathrm{tr}P\right\rangle _{\infty}\equiv\lim_{n\rightarrow
\infty}\left\langle n\right\vert \mathrm{tr}P\left\vert n\right\rangle
.\label{TrP}%
\end{equation}
The general configurations $P$ must be a vacuum at the boundary of the Hilbert
space and, as we shall see, $\left\langle \mathrm{tr}P\right\rangle _{\infty}$
takes the values $0,1,2.$ It can be easily verified that for the solutions
(\ref{sLee-P})(\ref{asLee-P})(\ref{sOSIK-P})(\ref{asOSIK-P}) $\left\langle
\mathrm{tr}P\right\rangle _{\infty}=1$, for (\ref{newS}) $\left\langle
\mathrm{tr}P\right\rangle _{\infty}=0$ , and for (\ref{newAS}) $\left\langle
\mathrm{tr}P\right\rangle _{\infty}=2$.\ 

The typical example of the vacuum configuration with zero energy,
\begin{equation}
P=\left(
\begin{array}
[c]{cc}%
1 & 0\\
0 & 0
\end{array}
\right)  ,
\end{equation}
corresponds to $\left\langle \mathrm{tr}P\right\rangle _{\infty}=1$
\footnote{We shall see that the general vacuum configurations with
$\left\langle \mathrm{tr}P\right\rangle _{\infty}=1$ are (\ref{vacP1}) and
(\ref{vacP2}).}. The vacua with $\left\langle \mathrm{tr}P\right\rangle
_{\infty}=0,2$ are respectively%
\begin{equation}
P=\left(
\begin{array}
[c]{cc}%
0 & 0\\
0 & 0
\end{array}
\right)
\end{equation}
and%
\begin{equation}
P=\left(
\begin{array}
[c]{cc}%
1 & 0\\
0 & 1
\end{array}
\right)  .
\end{equation}
The solitons can be considered to be the excited states from the respective
vacua. On the other hand, for the nonlinear sigma model on the commutative
space, we have
\begin{equation}
P=\frac{1}{2}+\frac{1}{2}\sum_{a=1}^{3}n^{a}\sigma^{a},
\end{equation}
where the space dependence is in $n^{a}$. Consequently, on the commutative
space, we always have
\begin{equation}
\;\left\langle \mathrm{tr}P\right\rangle _{\infty}\Rightarrow\;\lim
_{\left\vert x\right\vert \rightarrow\infty}\mathrm{tr}P=1,
\end{equation}
thus the configurations with $\left\langle \mathrm{tr}P\right\rangle _{\infty
}=0,2$ are characteristic of the nonlinear sigma model on the noncommutative
space. As we shall see, $\left\langle \mathrm{tr}P\right\rangle _{\infty}$ is
a conserved quantity against the continuous deformation of configuration with
finite energy. Consequently, configurations of the nonlinear sigma model are
classified in terms of the topological charge $Q$ and the value of
$\left\langle \mathrm{tr}P\right\rangle _{\infty}$. In what follows, we shall
show that $\left\langle \mathrm{tr}P\right\rangle _{\infty}$ is conserved
under the continuous deformations of the configurations.

We consider first the vacuum configurations with energy $E=0$ . Let us
parametrize $P$ as%
\begin{equation}
P=\left(
\begin{array}
[c]{cc}%
a & b\\
b^{\dagger} & c
\end{array}
\right)  .
\end{equation}
From%
\begin{equation}
P^{\dagger}=P,
\end{equation}
we have%
\begin{equation}
a^{\dagger}=a,\;c^{\dagger}=c. \label{AandC}%
\end{equation}
If we define%
\begin{equation}
A\equiv\left[  a,\bar{z}\right]  ,\;B\equiv\left[  b,\bar{z}\right]
,\;C\equiv\left[  b^{\dagger},\bar{z}\right]  ,\;D\equiv\left[  c,\bar
{z}\right]  ,
\end{equation}
the energy of the static configuration can be written as%
\begin{align}
E  &  =\frac{1}{\theta^{2}}\mathrm{Tr}_{\mathcal{H}}\left\{  \mathrm{tr}%
(\left[  z,P\right]  \left[  P,\bar{z}\right]  )\right\} \nonumber\\
&  =\frac{2\pi}{\theta}\sum\limits_{n=0}^{\infty}\left\langle n\right\vert
\left(  A^{\dagger}A+B^{\dagger}B+C^{\dagger}C+D^{\dagger}D\right)  \left\vert
n\right\rangle \nonumber\\
&  =0.
\end{align}
Consequently, for all $\left\vert n\right\rangle $\ we have%
\begin{equation}
\left\langle n\right\vert \left(  A^{\dagger}A+B^{\dagger}B+C^{\dagger
}C+D^{\dagger}D\right)  \left\vert n\right\rangle =0,
\end{equation}
from which it follows that%
\begin{equation}
A=B=C=D=0.
\end{equation}
Thus taking into account (\ref{AandC}), each component of $P$ is constant for
the configuration with $E=0$, and $P$ can be written as
\begin{equation}
P=\left(
\begin{array}
[c]{cc}%
a & b\\
\bar{b} & c
\end{array}
\right)  ,
\end{equation}
where $a,c$ are real numbers and $b$ is complex. Next from%
\begin{equation}
P^{2}=P,
\end{equation}
we have%
\begin{align}
a^{2}+b\bar{b}  &  =a,\nonumber\\
ab+bc  &  =b,\nonumber\\
\bar{b}b+c^{2}  &  =c.
\end{align}
Solving these we are lead to the two conditions given by%
\begin{equation}
a=\lambda_{\pm}\;,\ c=\lambda_{\pm}\ ,
\end{equation}
where%

\begin{equation}
\lambda_{\pm}\equiv\frac{1\pm\sqrt{1-4\left\vert b\right\vert ^{2}}}{2},
\label{condition1}%
\end{equation}
and
\begin{equation}
a+c=1\;\text{for}\;b\neq0. \label{condition2}%
\end{equation}
Possible vacuum configurations are classified into the following three types.
First, the configurations connected to
\begin{equation}
P=\left(
\begin{array}
[c]{cc}%
1 & 0\\
0 & 0
\end{array}
\right)  , \label{vacP0}%
\end{equation}
can be parametrized by the complex number $b$ $\left(  0\leq\left\vert
b\right\vert \leq\frac{1}{2}\right)  $ as
\begin{equation}
P=\left(
\begin{array}
[c]{cc}%
\lambda_{+} & b\\
\bar{b} & \lambda_{-}%
\end{array}
\right)  \label{vacP1}%
\end{equation}
and%
\begin{equation}
P=\left(
\begin{array}
[c]{cc}%
\lambda_{-} & b\\
\bar{b} & \lambda_{+}%
\end{array}
\right)  \ . \label{vacP2}%
\end{equation}
When we continuously deform the configuration in the region $0\leq\left\vert
b\right\vert \leq\frac{1}{2}$, $\left\langle \mathrm{tr}P\right\rangle
_{\infty}$ remains $1$. The other two types of vacuum configurations are
\begin{equation}
P=\left(
\begin{array}
[c]{cc}%
0 & 0\\
0 & 0
\end{array}
\right)  ,
\end{equation}
with $\left\langle \mathrm{tr}P\right\rangle _{\infty}=0$ and
\begin{equation}
P=\left(
\begin{array}
[c]{cc}%
1 & 0\\
0 & 1
\end{array}
\right)  ,
\end{equation}
with $\left\langle \mathrm{tr}P\right\rangle _{\infty}=2$. These two vacuum
configurations cannot be deformed keeping $E=0$ under the conditions
(\ref{condition1}) and (\ref{condition2}). Consequently, the vacua with
different values of $\left\langle \mathrm{tr}P\right\rangle _{\infty}$\ are
disconnected against the continuous deformations. Thus, $\left\langle
\mathrm{tr}P\right\rangle _{\infty}$ is a conserved quantity taking the values
$0,1,2$ under the continuous deformation keeping $E=0$.

We consider next the continuous deformation of the general configuration with
finite energy. In order to keep the energy%
\begin{equation}
E=\frac{2\pi}{\theta}\sum\limits_{n=0}^{\infty}\left\langle n\right\vert
\mathrm{tr}\left(  \left[  z,P\right]  \left[  P,\bar{z}\right]  \right)
\left\vert n\right\rangle \ ,
\end{equation}
finite for the static configuration $P$, the condition,
\begin{equation}
\lim_{n\rightarrow\infty}\left\langle n\right\vert \mathrm{tr}\left(  \left[
z,P\right]  \left[  P,\bar{z}\right]  \right)  \left\vert n\right\rangle =0\ ,
\end{equation}
is needed as the boundary condition. Consequently, general configuration $P$
must be a vacuum at the boundary of the Hilbert space ($\left\vert
n\right\rangle $ with $n\rightarrow\infty$), and thus $\left\langle
\mathrm{tr}P\right\rangle _{\infty}$\ takes the value $0$ or $1$ or $2$. As a
result, $\left\langle \mathrm{tr}P\right\rangle _{\infty}$ is a conserved
quantity and the configurations are classified by the topological charge
$Q=0,\pm1,\pm2,\cdots$\ and $\left\langle \mathrm{tr}P\right\rangle _{\infty
}=0,1,2$ .

\section{Summary}

On the commutative space, there exists a definite correspondence between the
configurations of the nonlinear sigma model and the configurations of $CP^{1}$
and both models are equivalent. We have seen, however, that on the
noncommutative space such a correspondence is destroyed. In fact, there exist
the BPS solitons (\ref{newS}) in the nonlinear sigma model that do not have
the counterpart in $CP^{1}$ model. On the other hand, the new BPS anti-soliton
solutions in the $CP^{1}$ model have been found in the noncommutative space
((\ref{newAScp}), general form is (\ref{newAScp-g})) that do not exist in the
commutative space.

We found that the configurations in the nonlinear sigma model is to be
classified not only by the topological charge $Q=0,\pm1,\pm2,\cdots$ but also
by $\left\langle \mathrm{tr}P\right\rangle _{\infty}=0,1,2.$ We have seen in
section 3, that for the configuration with $\left\langle \mathrm{tr}%
P\right\rangle _{\infty}=1$ there exist both solitons and anti-solitons. In
the case of $\left\langle \mathrm{tr}P\right\rangle _{\infty}=0$, only
solitons ($Q>0$) can exist, while in the case of $\left\langle \mathrm{tr}%
P\right\rangle _{\infty}=2$ only anti-solitons ($Q<0$) are confirmed. This
asymmetry deserves a further study.

Relations with the gauge theories and use of the solitons in the actual
physical problems are interesting topics to be investigated.

\bigskip

%\bibliographystyle{JHEP}
%\bibliography{refs2}

\providecommand{\href}[2]{#2}\begingroup\raggedright

\end{document}